\documentstyle[12pt,a4wide,epsf]{article}

\begin{document}
\begin{center}
{\large\bf 
Ultra slow-roll inflation demystified}

\medskip

{\large Konstantinos Dimopoulos}

\smallskip

{\em Consortium for Fundamental Physics}\\
{\em Physics Department, Lancaster University, Lancaster LA1 4YB, UK}

\smallskip

e-mail: {\tt k.dimopoulos1@lancaster.ac.uk}

\smallskip

\date{today}

\end{center}

\begin{abstract}
Ultra-slow-roll (USR) inflation is a new mode of inflation which corresponds to 
the occasions when the inflaton field must traverse an extremely flat part of 
the scalar potential, when the usual slow-roll (SR) fails. We investigate USR 
and obtain an estimate for how long it lasts, given the initial
kinetic density of the inflaton. We also find that, if the initial kinetic 
density is small enough, USR can be avoided and the usual SR treatment is valid.
This has important implications for inflection-point inflation.
\end{abstract}

\bigskip

\section{Introduction}

Cosmic inflation is an organic component of the concordance model of cosmology.
It is a period of exponential expansion in the early Universe, which determines
the initial conditions for the subsequent Hot Big Bang cosmology. In particular,
it makes the Universe spatially flat, large and uniform but also 
provides the necessary deviations from perfect uniformity in the form of the 
primordial curvature perturbation, which accounts for the eventual formation of 
the large scale structure. Typically, inflation is modelled through the 
inflationary paradigm, which suggests that the Universe undergoes inflation when
dominated by the potential density of a scalar field (inflaton). This potential 
density remains roughly constant during inflation. As a result, the generated 
curvature perturbation is almost scale-invariant, as suggested by observations. 
In order to keep the potential density roughly constant, the variation of the 
field must be very small throughout inflation. Because the inflaton's equation 
of motion is the same as a body rolling down a potential slope subject to 
friction, we need this roll to be slow for the inflaton, in field space, so as 
to keep the potential density roughly unchanged. Thus, in the inflationary 
paradigm, the inflaton undergoes slow-roll (SR) during inflation. Indeed, the 
latest CMB data favours single-field slow-roll inflation \cite{planck}. 

The SR solution is an attractor \cite{book} as long as the potential is flat 
enough to support it. However, it was recently realised that SR may end not 
only when the potential becomes steep and curved, as is for the end of 
inflation, but also when it suddenly becomes extremely flat, too flat for the 
regular SR assumptions to apply. In this case, 
the system engages in so-called ultra slow-roll (USR) inflation. This new mode 
of inflation has been hitherto unknown. It can have a profound impact on 
inflationary observables, so it must be taken into account. However, even 
though diagnosed, USR has not been fully understood, with most of its dynamics 
traced numerically. In this letter, we attempt to demystify USR and provide a 
conceptual understanding of its dynamics. Ignoring USR can lead to important
miscalculations of inflationary observables.

USR arises when the potential becomes
extremely flat, so much so that, SR would force the kinetic density of the field
to reduce faster than it would if the field were in free-fall 
\mbox{$\rho_{\rm kin}\equiv\frac12\dot\phi^2\propto a^{-6}$}, which of course 
cannot happen. Thus, the system departs from SR and the field engages in USR, 
during which the kinetic density decreases as in free-fall, until the system 
can get back to SR, when the decreasing $|\ddot\phi|$ catches up with the slope 
of the potential $|V'|$, or until inflation ends, e.g. by a phase transition. 
Note that, even though the slope is very small, we still have potential 
domination \mbox{$V>\rho_{\rm kin}$} so inflation continues. USR was first 
investigated in Ref.~\cite{kinney}, which was followed by 
Refs.~\cite{martin,sasaki} and recently by Ref.~\cite{germani}. In 
Refs.~\cite{kinney} and \cite{sasaki} a constant potential is assumed, which 
cannot exhibit SR. In Ref.~\cite{martin} it was shown that USR is not an 
attractor solution and the system departs from it as soon as the conditions 
which enforce USR allow it. But which conditions are these? 

In this letter we explore this question. To obtain an insight of the dynamics of
USR, we study USR in linear inflation and then generalise our findings for an
arbitrary inflation model. We particularly consider inflection-point inflation
because it can lead to USR. It is fair to say that the community seems little 
aware of USR, so the hope is that our treatment may be revealing of USR's 
nature.
This is a particularly acute problem in models of inflection-point inflation, 
where a region of USR exists around the inflection point. In USR this region is 
traversed in a moderate number of e-folds. However, were SR assumed, this 
number would grow substantially. As inflationary observables are determined by 
the correct number of e-folds, this can have profound implications on 
inflationary predictions and on the viability of inflection-point models. 

We use natural units, where \mbox{$c=\hbar=1$} and \mbox{$8\pi G=m_P^{-2}$},
with \mbox{$m_P=2.43\times 10^{18}\,$GeV} being the reduced Planck mass.

\section{Ultra-slow roll inflation}

To explore USR inflation, we will look closely at the Klein-Gordon equation of 
motion of the canonical homogeneous inflaton field $\phi$:
\begin{equation}
\ddot\phi+3H\dot\phi+V'=0\,,
\label{KG0}
\end{equation}
where $H\equiv\dot a/a$ is the Hubble parameter 
(with $a$ being the scale factor)
$V$ is the scalar potential and the dot \{prime\} denotes derivative with 
respect to the cosmic time \{the inflaton field\}. We name each term of the 
above as the acceleration, the friction and the slope term respectively. We 
also employ the flat Friedman equation during inflation, when the Universe is 
dominated by the inflaton field:
\begin{equation}
3H^2m_P^2=\frac12\dot\phi^2+V\,.
\label{fried}
\end{equation}

We define two slow-roll parameters
\begin{equation}
\epsilon\equiv-\dot H/H^2
\label{eps}
\end{equation}
and
\begin{equation}
\epsilon_2\equiv\frac{\dot\epsilon}{\epsilon H}
=-6-\frac{2V'}{H\dot\phi}+2\epsilon
=\frac{2\ddot\phi}{H\dot\phi}+2\epsilon\,,
\label{e2}
\end{equation}
where we have employed Eqs.~(\ref{KG0}) and (\ref{fried}). It is easy to show 
that 
\begin{equation}
\epsilon=\frac32(1+w)\,,
\label{ew}
\end{equation}
where $w$ is the barotropic parameter of the homogeneous inflaton field, 
given by
\begin{equation}
w=\frac{\rho_{\rm kin}-V}{\rho_{\rm kin}+V}\,,
\label{w}
\end{equation}
where \mbox{$\rho_{\rm kin}\equiv\frac12\dot\phi^2$}. For inflation we need
\mbox{$w<-\frac13$}, which means \mbox{$V>2\rho_{\rm kin}$}. From Eq.~(\ref{ew}),
we see that inflation (accelerated expansion) occurs when 
\mbox{$\epsilon<1$}.

Now, in the usual SR, the acceleration term in Eq.~(\ref{KG0}) is negligible, 
so the latter becomes
\begin{equation}
3H\dot\phi\simeq -V\,,
\label{SR}
\end{equation}
which shows that the friction term is locked to the slope term. In this case,
Eq.~(\ref{e2}) becomes
\begin{equation}
\epsilon_2=-2\eta+4\epsilon\,,
\label{e2SR}
\end{equation}
where the usual SR parameters are
\begin{equation}
\epsilon\simeq\epsilon_{\rm SR}\equiv\frac12 m_P^2\left(\frac{V'}{V}\right)^2
\quad{\rm and}\quad
\eta\equiv m_P^2\frac{V''}{V}\,.
\label{SRparam}
\end{equation}
During SR, $\epsilon,|\eta|\ll 1$, which means that \mbox{$|\epsilon_2|\ll 1$}.

However, if the potential suddenly becomes extremely flat, the slope term in 
Eq.~(\ref{KG0}) may reduce drastically, which means that it virtually 
disappears. The equation is then rendered
\begin{equation}
\ddot\phi+3H\dot\phi\simeq 0\,,
\label{USR}
\end{equation}
which shows that the friction term is now locked with the acceleration term.
In this case, Eq.~(\ref{e2}) becomes
\begin{equation}
\epsilon_2=-6+2\epsilon\,,
\label{e2USR}
\end{equation}
During inflation \mbox{$\epsilon<1$}, which means 
\mbox{$|\epsilon_2|\approx 6$}. Thus, if during inflation, the potential becomes
suddenly very flat, $|\epsilon_2|$, which is initially small grows to larger 
than unity, SR is applicable no-more and a period of USR begins.

Intuitively, one can understand this as follows. If we are in SR but the slope 
$|V'|$ reduces drastically, it initially drags with it the friction term, by
virtue of Eq.~(\ref{SR}). This decreases the value of $|\dot\phi|$, i.e. the 
kinetic density \mbox{$\rho_{\rm kin}=\frac12\dot\phi^2$}, but this value cannot 
decrease arbitrarily quickly. The fastest it can decrease is 
\mbox{$\rho_{\rm kin}\propto a^{-6}$}, which we call free-fall because it 
corresponds to a field with no potential density \mbox{$V=0$}, such that its 
equation of motion is Eq.~(\ref{USR}). Therefore, if the kinetic density of SR 
is forced (by the decreasing slope) to reduce faster than free-fall then the 
system breaks away from SR. In SR the acceleration term is negligible, because 
it is very small, compared to the friction and slope terms, which are locked 
together as shown in Eq.~(\ref{SR}). However, if the slope reduces 
drastically and drags the friction term with it, they both become small too and 
eventually comparable to the acceleration term. So all three terms in 
Eq.~(\ref{KG0}) are comparable. When this happens, the friction term changes 
allegiances and becomes locked with the acceleration term, resulting in USR.

Now, once in USR, the field becomes oblivious of the potential, as demonstrated 
by Eq.~(\ref{USR}). This is similar to the kination period of quintessential 
inflation models \cite{kination,mine} but there is a crucial difference. In 
kination, the Universe is dominated by $\rho_{\rm kin}$, while in USR inflation,
we still have potential domination and \mbox{$V>\rho_{\rm kin}$}. Being oblivious 
to the potential, the inflaton field can even climb up an ultra-shallow $V$
\cite{martin}. Indeed, when the system enters the USR regime, it ``flies over'' 
the flat patch of the potential,
sliding on its decreasing kinetic density. In that sense, the term ultra-SR is
actually a misnomer, because the field rolls faster than it would have done if 
SR were still applicable over the extremely flat region. 

Indeed, if $|V'|$ decreases to almost zero, 
so does $\epsilon_{\rm SR}$. In SR the number of elapsing e-folds is
\begin{equation}
\Delta N=
\frac{1}{m_P}\int_{\phi_1}^{\phi_2}\frac{{\rm d}\phi}{\sqrt{2\epsilon_{\rm SR}}}\,,
\end{equation}
which increases substantially if $\epsilon_{\rm SR}$ becomes extremely small.
In contrast, in USR $\epsilon$ does not decrease too much, so we have 
\mbox{$\epsilon_{\rm SR}\ll\epsilon<1$}. The number of elapsing e-folds is given 
in general by
\begin{equation}
\Delta N=-\int\frac{{\rm d}H}{\epsilon H}
\end{equation}
and in USR it can be much smaller compared to SR if 
\mbox{$\epsilon_{\rm SR}\ll\epsilon$}. 
Thus, when considering an inflation model that results in 
periods of USR, but only SR is assumed, there is a danger of overestimating the
number of e-folds it takes for the field to roll down. 

It is evident that USR depends on having substantial kinetic density, which 
cannot decrease faster than free-fall. However, if one begins inflation at the 
extremely flat region with very small kinetic density, then SR may be attained,
quickly, even immediately. Now, the initial conditions for inflation are 
shrouded by the no-hair theorem, which renders them academic, because all memory
is lost once the inflationary attractor is reached. Thus, provided inflation 
begins comfortably before the cosmological scales exit the horizon, the initial 
conditions of the inflaton field can be taken to correspond to kinetic density 
small enough to avoid USR despite an extremely flat scalar potential. This 
can rescue inflation models such as inflection-point inflation, which may have
problems with USR. To quantify how small the initial kinetic density needs to 
be, we first investigate linear inflation.

\section{Ultra-slow-roll in linear inflation and beyond}

We consider the inflation model:
\begin{equation}
V=V_0+M^3\phi\,,
\label{V}
\end{equation}
where $V_0$ is a constant density scale and $M$ is a mass scale. Then the 
Klein-Gordon Eq.~(\ref{KG0}) becomes:
\begin{equation}
\ddot\phi+3H_0\dot\phi+M^3=0\,,
\label{KG}
\end{equation}
where \mbox{$H_0^2\equiv V_0/3m_P^2$} and we assumed \mbox{$V_0\gg M^3\phi$}.
The above has the general solution
\begin{equation}
\dot\phi=Ce^{-3\Delta N}-\frac{M^3}{3H_0}\,,
\label{solu}
\end{equation}
where \mbox{$\Delta N=H_0 \Delta t$} is the elapsing e-folds 
and $C$ is a constant. We also find
\begin{equation}
\ddot\phi=-3H_0Ce^{-3\Delta N}\,.
\label{solu+}
\end{equation}

If initially ($\Delta N=0$) the velocity of the field is
$\dot\phi_0=0$, then $C=M^3/3H_0$ and the Klein-Gordon suggests 
that $\ddot\phi_0=-M^3$. Then, as time continues, the above suggest that the 
Klein-Gordon becomes
\begin{equation}
-M^3e^{-3\Delta N}+M^3(e^{-3\Delta N}-1)+M^3=0\,.
\end{equation}
Notice that, even though the friction term begins as zero it soon (in a single 
e-fold) dominates over the acceleration term and the slow-roll (SR) condition is
recovered, where $\ddot\phi$ is negligible and $V'$ is balanced by $3H\dot\phi$.
Thus if we start with zero velocity, we have SR immediately afterwards.

Now suppose that, originally $\dot\phi\neq 0$. If \mbox{$|C|\ll M^3/3H_0$} then
\mbox{$\dot\phi\simeq\dot\phi_0\simeq -M^3/3H_0$} (cf. Eq.~(\ref{solu})),
which means that the friction term is \mbox{$3H\dot\phi\simeq -M^3=V'$} and we 
have SR. Thus we always obtain immediately SR if \mbox{$|C|\leq M^3/3H_0$}.
If \mbox{$|C|\gg M^3/3H_0$} then
\mbox{$3H_0|\dot\phi_0|\simeq 3H_0|C|>M^3$}, which means that the friction term
initially dominates over the slope term and is balanced by the acceleration 
term, \mbox{$|\ddot\phi_0|=3H_0|C|$} according to Eq.~(\ref{solu+}). Thus, the 
Klein-Gordon is \mbox{$\ddot\phi+3H\dot\phi\simeq 0$} (cf. Eq.~(\ref{USR})), 
which gives rise to USR.
USR continues until \mbox{$3H_0|C|e^{-3\Delta N}=M^3$}, 
when all three terms in the Klein-Gordon become comparable. Afterwards, the 
friction term becomes \mbox{$3H\dot\phi\simeq -M^3$}, which counterbalances the 
slope term, while the acceleration term becomes negligible. Thus, we recover~SR.

Therefore, USR lasts
\begin{equation}
\Delta N_{\rm USR}=\frac13\ln\left(\frac{3H_0|C|}{M^3}\right)
=\frac13\ln\left(\frac{3H_0\sqrt{2\rho_{\rm kin}^0}}{M^3}\right),
\label{NUSR}
\end{equation}
where \mbox{$\rho_{\rm kin}^0\equiv\frac12\dot\phi_0^2$} is the initial kinetic 
density, which is \mbox{$\rho_{\rm kin}^0\simeq \frac12 C^2$} for large $|C|$.

All in all, we find that, to obtain a sizeable period of USR, we need
\begin{equation}
|C|\gg M^3/3H_0\quad\Leftrightarrow\quad
\rho_{\rm kin}^0
>\frac{M^6}{18H_0^2}.
\label{USRcondition}
\end{equation}
Otherwise, we have SR only. Note that, if \mbox{$M=0$} and the potential is 
exactly flat, SR is never recovered \cite{kinney}.

We may generalise the above for an arbitrary potential, as follows. At
extremely flat region of the potential we set $M^3\equiv V'(\phi_f)$ and 
$H_0^2=V(\phi_f)/3m_P^2$ and enforce the bound in Eq.~(\ref{USRcondition}),
where $\phi_f$ corresponds to the flattest part of the potential. Thus, to 
avoid USR we need
\begin{equation}
\rho_{\rm kin}(\phi_f)\equiv
\frac12\dot\phi^2_f\leq\left.\frac{(V')^2m_P^2}{6V}\right|_{\phi_f}=
\frac13\epsilon_{\rm SR}(\phi_f)V(\phi_f)\,.
\label{bound}
\end{equation}
The above makes sense, because the kinetic density in SR is
\begin{equation}
\rho_{\rm kin}^{\rm SR}=\frac12\left(\frac{V'}{3H}\right)^2=
\frac13\epsilon_{\rm SR}V\,,
\end{equation}
where we used Eqs.~(\ref{SR}) and (\ref{SRparam}). Thus, the bound in 
Eq.~(\ref{bound}) really requests that the kinetic density in the flat patch be
at most the one corresponding to SR:
\mbox{$\rho_{\rm kin}(\phi_f)\leq\rho_{\rm kin}^{\rm SR}$}. 
This makes sense because if one has 
kinetic density in excess of $\rho_{\rm kin}^{\rm SR}$, the friction term in 
Eq.~(\ref{KG0}) cannot be balanced by the slope term and we have USR.

In view of the above, we can also recast Eq.~(\ref{NUSR}) as
\begin{equation}
\Delta N_{\rm USR}=
\frac16\ln\left(\frac{3\rho_{\rm kin}^0}{\epsilon_{\rm SR}V}\right)\,,
\end{equation}
where we used the potentially dominated Friedman equation.




\section{Ultra-slow-roll in inflection-point inflation}

We now focus on inflection-point inflation, which may feature USR. 
Inflection point inflation corresponds to the case of a flat step on the 
otherwise steep potential wall. This step is formed because of opposing 
terms in the potential which almost cancel each other. There are many model 
realisations, most notably A-term inflation \cite{A-term}, MSSM inflation
\cite{MSSM} and many others \cite{inflection}. However, in the vast majority of 
these works the USR phase has not been considered, which may cast doubt on some 
of their findings.

To avoid the USR period, one only needs to assume that the initial kinetic 
density is small enough according to the bound in Eq.~(\ref{bound}), where 
$\phi_f$ now corresponds to the inflection point, which is the flattest part 
of the potential plateau. This can be understood as follows. 
The potential for inflection-point inflation can be crudely approximated by 
three consecutive segments of linear potential.
Inflation only takes place along the flattest segment, and it is similar to 
linear inflation.

While rolling from large values of $\phi$ to small, when the field reaches 
the flat segment
then there is an abrupt reduction in $|V'|$. Because the friction term in the 
Klein-Gordon was at least as large as the slope term before reaching 
the flat segment
(i.e. we had SR or free-fall), after
wards, the friction term cannot be balanced by the (substantially 
reduced) slope term. Thus, the acceleration term rushes to balance it and we 
have USR. 

Now, during USR, we have \mbox{$\rho_{\rm kin}\propto a^{-6}$} so that
\begin{equation}
\dot\phi\ddot\phi=\dot\rho_{\rm kin}=-6H\rho_{\rm kin}\propto a^{-6},
\end{equation}
where we took \mbox{$H\simeq\,$constant}. Because 
\mbox{$|\dot\phi|=\sqrt{2\rho_{\rm kin}}\propto a^{-3}$}, the above suggests that
\mbox{$|\ddot\phi|\propto a^{-3}$}. After crossing the inflection point, though,
the slope of the potential begins to increase, while the acceleration decreases,
as we have seen. At some point, they meet each other and then the friction term 
changes allegiances and becomes locked with the slope term, so that SR is 
recovered. 

But what if the evolution of the field had already begun at the flat patch? 
Then, provided the kinetic density is small enough, one can immediately have SR 
inflation \cite{antonio}. The bound in Eq.~(\ref{bound}) is a  conservative 
estimate on the maximum kinetic density because the slope at the inflection 
point is smaller than at the rest of the plateau.

\section{Quantum diffusion}

Now, we briefly discuss quantum diffusion. 
If the potential is extremely flat, quantum fluctuations of the field may 
dominate its variation. The quantum variation of the field per Hubble time 
\mbox{$\delta t=H^{-1}$} is typically given by the Hawking temperature in de 
Sitter space \mbox{$\delta\phi=H/2\pi$}. Thus, the kinetic density of quantum 
fluctuations is 
\begin{equation}
\rho_{\rm kin}^{\rm diff}=
\frac12\left(\frac{\delta\phi}{\delta t}\right)^2= \frac{H^4}{8\pi^4}\,.
\end{equation}
This should be interpreted
as the lowest value the kinetic density can have. If 
\mbox{$|V'|<\frac{3}{2\pi}H^3$} then quantum diffusion overwhelms SR%
\footnote{because $|\dot\phi|=|V'|/3H<H^2/2\pi=\frac{\delta\phi}{\delta t}$}, 
so that the number of USR e-folds is
\begin{equation}
\Delta N_{\rm USR}=
\frac16\ln\left(\frac{\rho_{\rm kin}^0}{\rho_{\rm kin}^{\rm diff}}\right)=
\frac13\ln\left(6\pi\,\frac{\sqrt{2\rho_{\rm kin}^0}\,m_P^2}{V}\right),
\end{equation}
where 
we used \mbox{$\rho_{\rm kin}\propto a^{-6}$} in USR and the potential dominated 
Friedman equation. 

\section{Perturbations}

We now comment briefly on the curvature perturbation during USR inflation. This
has been studied extensively in Ref.~\cite{sasaki}. Here we note that,
during USR there is a spike in the curvature perturbation, which may 
potentially lead to the copious production of primordial black holes, that can 
substantially contribute to the dark matter in the Universe \cite{germani,pbh}.
This can be understood as follows. 

For the spectrum of the curvature perturbation we have
\begin{equation}
\sqrt{\cal P}=\frac{H^2}{2\pi\dot\phi}\;\Rightarrow\;
{\cal P}=\frac{H^2}{8\pi^2 m_P^2\epsilon}\,,
\label{P}
\end{equation}
where we used that \mbox{$2m_P^2\dot H=-\dot\phi^2$} and Eq.~(\ref{eps}).

In SR inflation the variation of \mbox{$\epsilon=\epsilon_{\rm SR}$} is very 
small, so $\cal P$ remains roughly constant, which corresponds to an almost
scale-invariant spectrum of perturbations. Indeed, the variation of $\epsilon$
is traced by \mbox{$\epsilon_2\equiv\dot\epsilon/\epsilon H$} (c.f. 
Eq.~(\ref{e2})). In SR, Eq.~(\ref{e2SR}) suggests that 
\mbox{$|\epsilon_2|=|4\epsilon-2\eta|\ll 1$}. 

Things are different during USR, though. Because 
\mbox{$\epsilon=\frac32\dot\phi^2/V$}, where \mbox{$V\simeq 3m_P^2H^2$} and 
\mbox{$\dot\phi^2=2\rho_{\rm kin}\propto a^{-6}$} during USR inflation, we find 
\mbox{$\epsilon\propto a^{-6}\propto e^{-6\Delta N}$}, where $\Delta N$ is the 
elapsing USR e-folds. Thus, we obtain that \mbox{${\cal P}\propto e^{6\Delta N}$}
and the curvature perturbation grows exponentially during USR inflation.%
\footnote{Note that $\cal P$ must be evaluated at the end of USR and not at 
horizon exit \cite{kinney,sasaki}.}

Note, though, that, were USR not considered, according to Eq.~(\ref{P}), the 
usual SR would lead to an even more dramatic increase in $\cal P$ 
because \mbox{$\epsilon_{\rm SR}\ll\epsilon$} when traversing an extremely flat 
patch of the potential. This, however, does not happen as the field 
``overshoots'' the flat patch \cite{germani} ``surfing'' on its decreasing 
kinetic density.

\section{Conclusions}

In conclusion, we have investigated ultra-slow-roll (USR) inflation, which may 
take place when the inflationary potential becomes extremely flat. We have 
showed that this is a temporary phase of inflation, not an attractor, and 
obtained an estimate of how many e-folds it lasts, depending on 
the initial kinetic density of the inflaton field. We have discussed how the 
field can depart from the usual slow-roll (SR) when crossing an extremely flat 
patch in the scalar potential. SR would force the field to spend a lot of time
traversing the flat patch. Instead, the field ``glosses over'' the flat 
patch in a moderate number of e-folds. Because the number of e-folds is of 
paramount importance when calculating inflationary observables, we argued that 
USR has to be taken into account, when necessary. In particular, we looked into 
inflection-point inflation, which exhibits a flat patch near the inflection 
point in the potential, that may give rise to USR. Models which do not take this
into account are in danger of miscalculating the values of inflationary 
observables. However, this danger can be averted if one assumes that the field 
begins its evolution already on the flat patch (e.g. near the inflection point)
with small initial kinetic density. We obtained a conservative bound on the 
initial kinetic density of the field, which manages to avoid USR inflation and
render the SR treatment valid.

\paragraph*{Acknowledgements}

My research is supported (in part) by
the Lancaster-Manchester-Sheffield Consortium for Fundamental Physics under 
STFC grant: ST/L000520/1. I would like to thank C.~Germani and C.~Owen for 
discussions.

\end{document}